\documentclass[11pt]{article}

\usepackage[preprint]{acl}

\usepackage{times}
\usepackage{latexsym}

\usepackage[T1]{fontenc}

\usepackage[utf8]{inputenc}

\usepackage{microtype}

\usepackage{inconsolata}

\usepackage{graphicx}
\usepackage{booktabs}
\usepackage{tabularx}

\title{Communication Enhances LLMs' Stability in Strategic Thinking}

\author{Nunzio Lore \\
  Network Science Institute\\ Northeastern University \\ 
  Boston, MA 02115 \\ United States \\
  \texttt{lora.n@northeastern.edu} \\\And 
    Babak Heydari \\
 Department of Mechanical and Industrial Engineering,\\ Institute of Experiential AI \\ Network Science Institute \\
 Northeastern University, Boston, MA 02115, \\ United States \\
  \texttt{b.heydari@northeastern.edu} \\}

\begin{document}
\maketitle
\begin{abstract}
Large Language Models (LLMs) often exhibit pronounced context-dependent variability that undermines predictable multi-agent behavior in tasks requiring strategic thinking. Focusing on models that range from 7 to 9 billion parameters in size engaged in a ten-round repeated Prisoner’s Dilemma, we evaluate whether short, costless pre-play messages emulating the cheap-talk paradigm affect strategic stability. Our analysis uses simulation-level bootstrap resampling and nonparametric inference to compare cooperation trajectories fitted with LOWESS regression across both the messaging and the no-messaging condition. We demonstrate consistent reductions in trajectory noise across a majority of the model–context pairings being studied. The stabilizing effect persists across multiple prompt variants and decoding regimes, though its magnitude depends on model choice and contextual framing, with models displaying higher baseline volatility gaining the most.  While communication rarely produces harmful instability, we document a few context-specific exceptions and identify the limited domains in which communication harms stability. These findings position cheap-talk style communication as a low-cost, practical tool for improving the predictability and reliability of strategic behavior in multi-agent LLM systems.
\end{abstract}

\section{Introduction}

Large language models (LLMs) are increasingly studied and deployed as autonomous or delegated agents across a wide range of environments within the agentic AI paradigm \cite{acharya2025agentic, shavit2023practices, sapkota2025ai}. While this technology holds considerable promise, it is also subject to critical failure modes \cite{li2024survey, han2024llm}. Among these, strategic misalignment and behavioral instability have emerged as central concerns in multi-agent systems \cite{cemri2025multi}. Instability can arise not only from sampling noise, temperature settings, or hardware-level nondeterminism, but also from the pronounced context-dependence of modern LLMs \cite{lore2024strategic}. As these systems transition from theoretical possibility to active deployment and empirical investigation, strategic stability becomes vital, both for reliable operation \cite{novikova2025consistency} and for meeting human expectations regarding the behavior of agentic systems \cite{barak2025humans}. 

Concerns about instability extend beyond issues of reproducibility. In strategic environments that mirror the incentive misalignment characteristic of social dilemmas, even minor stochastic fluctuations can propagate into qualitatively and quantitatively distinct behavioral trajectories. Recent work shows that LLMs exhibit substantial variability in complex reasoning tasks \cite{wang2025assessing}, with subtle changes in framing or decoding parameters yielding divergent strategies \cite{song2024good, yang2025alignment}. When deployed in multi-agent contexts such as negotiation, coordination, or repeated games, this variability can degrade both individual performance and system-level dynamics.

We posit that behavioral stability is a functional prerequisite for strategic alignment. In multi-agent systems, the goal is often to steer agents toward high-utility outcomes (e.g., cooperation). However, without a predictable behavioral trajectory, attempts to optimize for utility are indistinguishable from stochastic drift. Before agents can be aligned to be "good," they must first be reliable. Therefore, this study prioritizes the reduction of trajectory variance over the maximization of payoff, treating stability as the foundational substrate upon which future utility-maximizing interventions can be reliably designed \cite{wolf2023fundamental}.

In this paper, we show that pre-play communication stabilizes strategic behavior in LLM agents. Allowing agents to exchange costless, non-binding messages before acting reduces stochastic drift and yields more consistent behavioral trajectories. Importantly, this stabilization operates as a general mechanism: communication constrains randomness while still permitting frame-dependent equilibria to emerge. Our approach builds on empirical work examining LLM behavior across contextual framings \cite{lore2024large, lore2024strategic}. We find clear departures from both classical cheap-talk predictions and human experimental benchmarks. Whereas game theory predicts babbling equilibria and human subjects often overshare strategic information, LLMs occupy a distinct behavioral regime. Characterizing this regime -- and identifying mechanisms that stabilize it -- is essential for the dependable deployment of multi-agent LLM systems. 

Our analysis focuses on the 7B-9B parameter regime, a choice motivated by three converging factors that position small-scale models as a critical domain for multi-agent research. First, from a deployment perspective, resource constraints in decentralized systems--such as edge computing swarms or high-frequency trading agents--necessitate lightweight architectures, making the stability of this class a prerequisite for scalable agentic AI. Second, theoretical efforts toward model ``miniaturization'' and distillation suggest that future compact models will increasingly inherit the capabilities of larger predecessors; understanding emergent behaviors in this size range thus serves as a bellwether for the next generation of efficient models. Third, smaller models generally exhibit lower strategic sophistication and higher susceptibility to noise than their larger counterparts \cite{zhang2024llm, baldwin2025impact}. By targeting this ``high-risk'' demographic, we subject our stabilization mechanism to the most rigorous possible testbed: if communication can stabilize these inherently volatile architectures, it offers a robust baseline for clearer strategic reasoning in more capable models.

We conduct a battery of simulations across four models, three prompt variations, and six contexts. In each simulation, two LLM agents play a ten-round repeated Prisoner’s Dilemma. We extract the average cooperation pattern and quantify its stability using locally weighted scatterplot smoothing (LOWESS), comparing the root mean squared error (RMSE) of the smoothed trajectories under messaging and no-messaging conditions. We find that pre-play communication produces smoother trends and lower RMSE, indicating greater stability. This stabilizing effect is not uniform: it depends on the interaction between model training, prompt formulation, and contextual framing. As an additional robustness check, we replicate the analysis with all models' temperatures set to 0. We further extend our analysis to network interaction, examining both the general case and cases where agents face communication constraints. We find that in both the zero-temperature and network settings, the effectiveness of messaging is mediated entirely by model choice, implying that innate strategic instability may not be mitigated simply by shifting the contextual framing. 

The remainder of the paper is organized as follows. Section 2 reviews the relevant literature. Section 3  details our methods and model selection, Section 4 presents our results, and Section 5 concludes. Sections 6 and 7 are reserved for our discussion on limitations and for our Ethics Statement, respectively. 

\section{Literature Review}

\subsection{Strategic Decision Making in LLMs}

Embedding true strategic competence, defined by a principled mechanism to evaluate the trade-offs of different responses and choices in an interactive decision making environment remains a core open problem for LLMs \cite{zhu2025gtalign}. Despite initially being postulated as a promising avenue to study humans "in silico", \cite{horton2023large}, recent studies indicate that when placed in adversarial or complex game-theoretic contexts, these models frequently exhibit bounded rationality and a reliance on heuristics \cite{sun2025game}. This bounded rationality is not perfectly humanlike, and often diverges from predictions of rational choice or behavioral models \cite{kitadai2023toward, zhang2024llm, guo2023gpt, mei2024turing, fontana2025nicer}. LLMs also struggle with belief refinement \cite{fan2024can} and from rigidity in applying humanlike strategies \cite{zheng2025beyond}. It has been argued that simulation of strategic human behavior through LLM is hampered by factors such as drawbacks in simulation design \cite{wang2025limits} and unobservable decisions made by researchers during the alignment phase \cite{willis2025will}. Given their widespread adoption, several studies have been conducted on how LLM system level performance can be improved and tuned towards prosocial goals \cite{justus2025llms, chen2024instigating, tran2025multi, wang2025beyond, piatti2024cooperate}. 

\subsection{Stability} 

Instability in the outputs of large language models (LLMs) has been attributed to a complex interplay of factors, including the inherent stochasticity of probabilistic sampling \cite{atil2024llm}, hardware and computational constraints \cite{atil2024non}, as well as variations introduced during pre-training and system design \cite{wang2025assessing, roy2025interpreting}. Prompt framing and contextual cues further modulate this variability, often amplifying differences across runs or deployments \cite{huang2024far, mozikov2024eai, errica2025did, sclar2023quantifying}. Notably, a growing body of evidence suggests that larger, more advanced, and ostensibly more capable models do not necessarily exhibit greater stability than their smaller predecessors \cite{zhou2024larger, yu2025performance}. As the deployment of LLMs accelerates across domains, this instability raises particular concerns for high-stakes applications that demand consistent and reproducible reasoning \cite{carandang2025llms, blair2025llms, purushothama2025not, gallagher2024assessing}. To address these challenges, a range of benchmarking frameworks have been developed and proposed to systematically assess and quantify model reliability \cite{anghel2025diagnosing, nalbandyan2025score, jang2025rcscore}. These efforts underscore the growing recognition that single-run performance metrics alone are insufficient for evaluating real-world robustness. In particular, \cite{li2025firm} show that in multi-turn interactions, LLMs often exhibit progressive degradation in stability and consistency, ultimately undermining their overall trustworthiness. Furthermore, the pronounced context-dependence observed in strategic decision-making constitutes a critical dimension of LLM instability in psychological and behavioral question-answering, where subtle variations in context, phrasing, or framing can lead to markedly divergent responses \cite{kovavc2024stick}.

\subsection{Strategic Communication}

The problem of strategic communication is pervasive across economic and social interactions and has been examined within game theory under a wide range of modeling assumptions \cite{spence1978job, akerlof1978market, grossman1981informational, milgrom1981good, schelling1980strategy, kamenica2011bayesian}. Most closely related to the setting modeled in this paper is the framework of cheap talk \cite{crawford1982strategic, farrell1996cheap}: communication that is both costless and non-binding, such that messages neither directly affect player payoffs nor carry enforceable truth requirements. Classical economic theory predicts “babbling” as the only equilibrium outcome: when incentives diverge, rational players are expected to exchange non-informative messages to preserve strategic ambiguity. Subsequent theoretical and experimental work has shown that repeated interaction can partially mitigate this tension, allowing for the gradual emergence of informative communication through mechanisms such as reputation formation and long-term payoff optimization \cite{golosov2014dynamic, blume2020strategic, wilson1997liar, lima2024infinitely, arechar2017m}. Yet, field experiments reveal that even in one-shot settings, human senders tend to overcommunicate, providing messages whose informational content exceeds equilibrium predictions \cite{cai2006overcommunication, sanchez2007experimental, lafky2022preferences, wang2010pinocchio}. The converse pattern, credulity, arises when receivers place excessive trust in messages despite being aware of the sender’s incentives, causing them to overweight biased information relative to equilibrium expectations \cite{li2022we, kartik2007credulity}. Other studies, by contrast, document an opposing tendency known as overcautiousness, in which receivers underweight signals because of risk aversion or generalized distrust \cite{li2016cheap}. These behavioral asymmetries are particularly relevant to the study of strategic reasoning in LLMs, given the ongoing debate over whether such models reproduce, amplify, or attenuate characteristic human biases in communication and strategic decision making \cite{tjuatja2024llms, geva2025llms}.

\begin{figure*}[!htbp]
    \centering
    \includegraphics[width=0.75\linewidth]{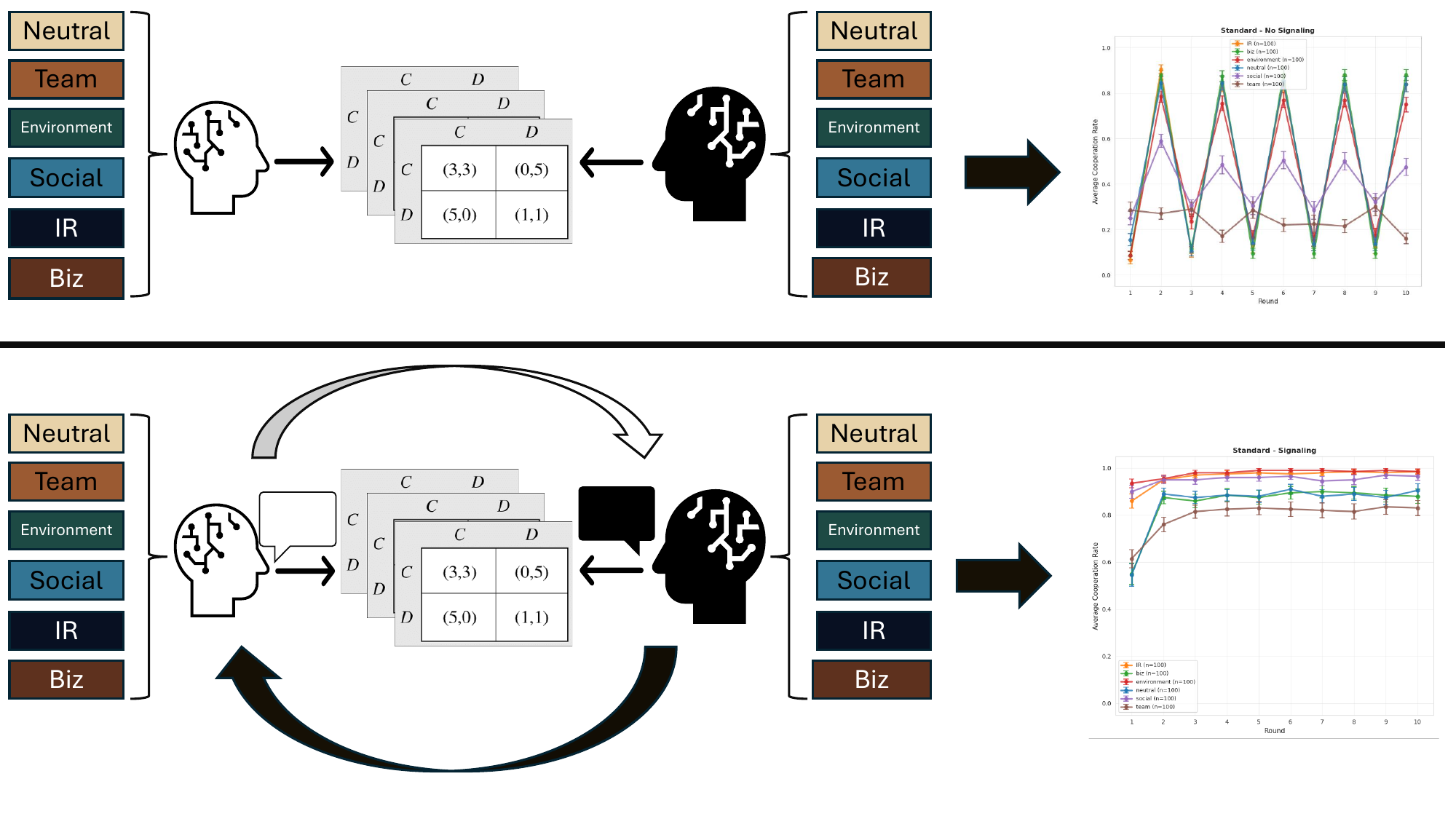}
    \caption{Overview of the general framework of this paper. Agents are presented with the payoff structure of the Prisoner's Dilemma and informed that they will play the game for multiple rounds, without ever specifying the time horizon. The prompt frames the interaction according to one of 5 + 1 social contexts, one of which (neutral) is used as a baseline. In the messaging treatment, agents can exchange a short one-sentence message before selecting their action each round. The cooperation trajectory for each model-context pair is then separately analyzed for each treatment using a bootstrapped LOWESS regression and calculating the RMSE of the fit. We study the average difference in RMSE between the no-messaging vs. messaging treatment, construct the empirical 95\% confidence intervals for the difference, and assess statistical significance using the exclusion principle.}
    \label{framework}
\end{figure*}

\section{Methods}

\subsection{Experimental Design}

We tested our conceptual framework using the iterated Prisoner's Dilemma (IPD) games across five contextual framings, plus a neutral context. We selected the IPD not merely for its canonical status, but because its infinite space of equilibria (per the Folk Theorem) maximizes the potential for strategic drift. Unlike coordination games (e.g., Stag Hunt) where dynamics often collapse quickly into basins of attraction, the IPD sustains a high tension between individual rationality and collective optimality. This structural feature prevents the system from settling trivially into a stable state, thereby making it the most rigorous testbed for evaluating whether communication can artificially induce stability in an inherently volatile environment. We never refer to the Prisoner's Dilemma explicitly, instead we merely provide the game's payoff structure. This is done to prevent the models from falling back on their pre-training knowledge, instead requiring the models to leverage their autoformalization capabilities.

For each model, we ran 100 independent simulations of ten-round games per context, a sample size chosen to reliably distinguish signal from stochastic noise. Agents are never made aware of the finite-time horizon. In the communication condition, agents exchanged one-sentence free-form messages before each decision. Agents are explicitly informed that messages are non-binding and instructed to interpret coplayer statements cautiously. In the control condition, agents acted based solely on play history. 

We distribute the context of the interaction together with rules of the game in the system-level prompt. This ensures adherence to the assigned roles and allows us to save tokens that would clutter the agents' context windows by distributing general level instructions only once. In each round, using user-level prompt, we inform each agent of their past round performance, ask them to send a message (under the messaging treatment), and then prompt the agents to select their action after receiving a message from their coplayer(s) (if any). 

To quantify the stability of strategic patterns, we developed a measure based on deviations from smooth trajectories. For each condition, which we define as the collection of simulations performed using a certain model, given a certain treatment and a context, we extracted individual agent decisions across all 10 rounds, yielding approximately 200 agent-level trajectories per condition from 100 simulations. We computed the average cooperation rate for each round across all agents within a condition, producing a 10-point time series of population-level cooperation rates.

We applied locally weighted scatterplot smoothing \cite{cleveland1979robust} with a bandwidth parameter of 0.4 to each condition-level time series. LOWESS is a non-parametric regression technique that fits locally weighted polynomial models, adapting flexibly to any smooth functional form, whether linear, monotonic, curved, or plateau-like, without imposing parametric assumptions. This approach captures the underlying strategic trajectory while remaining agnostic to its specific shape. We then calculated the root mean squared error (RMSE) between the observed round-level averages and the LOWESS-fitted values. Low RMSE indicates that population cooperation follows a predictable, smooth trajectory (high stability), while high RMSE indicates erratic fluctuations around the trend (low stability). This metric thus operationalizes strategic stability as the consistency with which population behavior adheres to a smooth developmental pattern. In this setting, RMSE ranges from 0, indicating a perfect fit, to 1, indicating maximal deviation from the smoothed trajectory.

To test whether messaging reduced RMSE (or, equivalently, increased stability), we employed a bootstrap procedure that properly accounts for the nested data structure. Since agents within the same simulation interact strategically and are therefore non-independent, we resampled at the simulation level: for each of 10,000 bootstrap iterations, we drew simulations with replacement (preserving both agents from each selected simulation), recalculated condition-level round averages, refit LOWESS curves, and computed bootstrap RMSE values for both no-messaging and messaging conditions. For each bootstrap iteration, we calculated the difference in RMSE (no-messaging minus messaging), yielding an empirical distribution of 10,000 difference values. We constructed 95\% confidence intervals and assessed statistical significance using the confidence interval exclusion criterion. This approach provides non-parametric inference that does not assume normality of the bootstrap distribution and correctly accounts for within-simulation dependencies.

To address potential inflation of Type I error from multiple testing, we conducted two binomial tests as omnibus assessments of the overall pattern. First, we tested whether the number of significant results (alpha = 0.05) exceeded the 1.2 expected by chance under the null hypothesis of no effect. Second, we tested whether the proportion of positive differences (indicating that messaging reduces RMSE) exceeded 50\%, which would be expected if messaging had no systematic effect. These tests provide population-level evidence that the observed pattern is unlikely to arise from chance variation across multiple comparisons. 

All simulations used \texttt{ollama} with temperature 0.8 (with the exception of a robustness check run at a temperature of 0 in the dyadic case). Contrary to most other papers in the field of LLM stability, we use a moderately high sampling temperature throughout the experiments. This choice reflects the study’s focus on the stability of strategic behavior rather than deterministic performance. Temperature controls the entropy of the model’s output distribution; higher values encourage exploration of alternative actions, analogous to mixed strategies in classical game theory. Sampling at T = 0.8 therefore exposes the model’s underlying policy distribution rather than its single most probable action, allowing us to assess whether strategic patterns evolve smoothly or erratically across repeated interactions. Importantly, this temperature range (0.7–0.9) is standard in prior work examining behavioral diversity and multi-agent dynamics in language models \cite{aher2023using, lore2024strategic}. In this sense, high-temperature sampling is not a source of uncontrolled noise but a deliberate probe of policy stability under symmetrical limiting conditions.

In the networked case, we reduce our number of simulations per model from 100 to 10 and deploy 50 agents per network. In order to lighten the computational load, we ask agents to specify all of their actions for all of their neighbors in the same message each round, leveraging a specific format that allows for easy parsing. We also modify the dynamics of messaging: instead of allowing one message per neighbor, we ask agents to commit to one message for the entire neighborhood. 

We leveraged three distinct network topologies generated via the \verb|networkx| library. First, we employed Erdős-Rényi (ER) random graphs with an edge creation probability of $p=0.1$. Second, we generated power law networks using the Holme-Kim algorithm to capture heavy-tailed degree. We set the attachment parameter $m=4$ to ensure a minimum degree of 4, and the triangle formation probability to $0.1$ to induce local clustering while preserving the dominance of central hubs. Finally, we constructed Core-Periphery structures using a Stochastic Block Model (SBM). The population was partitioned into a dense core ($20\%$ of nodes) and a sparse periphery ($80\%$). Connection probabilities were calibrated to enforce hierarchy while ensuring connectivity: within-core density was set to $\approx 0.56$, cross-group density to $\approx 0.10$, and within-periphery density to $\approx 0.06$.

\begin{table*}[t]
\centering
\scriptsize
\renewcommand{\arraystretch}{0.85}
\begin{tabularx}{\textwidth}{llcccccc}
\toprule
\textbf{Model} & \textbf{Context} & \textbf{RMSE (No Messaging)} & \textbf{RMSE (Messaging)} & \textbf{Difference} & \textbf{95\% CI Lower} & \textbf{95\% CI Upper} \\
\midrule
\textbf{QWEN 2.5 7b} & neutral      & 0.3742 & 0.0577 & \textbf{0.3165}\textsuperscript{†} & 0.2566 & 0.3949 \\
               & biz          & 0.4082 & 0.0045 & \textbf{0.4037}\textsuperscript{†} & 0.3129 & 0.4383 \\
               & environment  & 0.3102 & 0.0043 & \textbf{0.3059}\textsuperscript{†} & 0.2495 & 0.3598 \\
               & social       & 0.1154 & 0.0088 & \textbf{0.1065}\textsuperscript{†} & 0.0599 & 0.1636 \\
               & team         & 0.0432 & 0.0148 & 0.0284 & -0.0019 & 0.0608 \\
               & IR           & 0.3850 & 0.0112 & \textbf{0.3738}\textsuperscript{†} & 0.3114 & 0.4297 \\
\midrule
\textbf{FALCON 3 7b} & neutral      & 0.0515 & 0.0323 & 0.0192 & -0.0136 & 0.0615 \\
                 & biz          & 0.0691 & 0.0183 & \textbf{0.0509}\textsuperscript{†} & 0.0094 & 0.0875 \\
                 & environment  & 0.0598 & 0.0180 & \textbf{0.0418}\textsuperscript{†} & 0.0005 & 0.0744 \\
                 & social       & 0.0142 & 0.0308 & -0.0166 & -0.0425 & 0.0120 \\
                 & team         & 0.0228 & 0.0266 & -0.0037 & -0.0321 & 0.0259 \\
                 & IR           & 0.0531 & 0.0138 & \textbf{0.0393}\textsuperscript{†} & 0.0011 & 0.0757 \\
\midrule
\textbf{GRANITE 3.3 8b} & neutral      & 0.1882 & 0.0443 & \textbf{0.1438}\textsuperscript{†} & 0.0831 & 0.2000 \\
                  & biz          & 0.3009 & 0.0838 & \textbf{0.2171}\textsuperscript{†} & 0.1523 & 0.2739 \\
                  & environment  & 0.3701 & 0.0457 & \textbf{0.3244}\textsuperscript{†} & 0.2658 & 0.3757 \\
                  & social       & 0.3077 & 0.0320 & \textbf{0.2757}\textsuperscript{†} & 0.2042 & 0.3262 \\
                  & team         & 0.3283 & 0.0163 & \textbf{0.3120}\textsuperscript{†} & 0.2477 & 0.3582 \\
                  & IR           & 0.1890 & 0.0243 & \textbf{0.1647}\textsuperscript{†} & 0.1017 & 0.2243 \\
\midrule
\textbf{GEMMA 2 9b} & neutral      & 0.0753 & 0.0091 & \textbf{0.0662}\textsuperscript{†} & 0.0195 & 0.1163 \\
                & biz          & 0.0203 & 0.0039 & \textbf{0.0164}\textsuperscript{†} & 0.0054 & 0.0407 \\
                & environment  & 0.0141 & 0.0450 & -0.0308 & -0.0478 & 0.0144 \\
                & social       & 0.0976 & 0.0322 & \textbf{0.0654}\textsuperscript{†} & 0.0079 & 0.1016 \\
                & team         & 0.0257 & 0.0382 & -0.0125 & -0.0466 & 0.0195 \\
                & IR           & 0.0313 & 0.0252 & 0.0061 & -0.0153 & 0.0417 \\
\midrule
\multicolumn{7}{l}{\textbf{Multiple-comparison adjustment}} \\
\multicolumn{3}{l}{Excess significant results} & \multicolumn{2}{l}{17/24 significant (expected 1.2)} & \multicolumn{2}{l}{$p < 0.001$} \\
\multicolumn{3}{l}{Directional consistency} & \multicolumn{2}{l}{20/24 positive (83.3\%)} & \multicolumn{2}{l}{$p < 0.001$} \\
\bottomrule
\end{tabularx}
\caption{Non-parametric bootstrap analysis of messaging effect on RMSE of LOWESS fit across all models and contexts. RMSE differences between no-messaging and messaging conditions were evaluated using bootstrap resampling at the simulation level (10,000 iterations), preserving within-simulation agent dependencies. This non-parametric approach constructs confidence intervals from the empirical distribution of differences without assuming normality. A positive difference indicates a reduction in RMSE in the messaging treatment, and vice-versa a negative value indicates an increase in RMSE under messaging. Significant effects (95\% CI excluding zero) are marked with \textsuperscript{†}. To address potential inflation of Type I error from multiple testing, we conducted two binomial tests as omnibus assessments of the overall pattern. First, we tested whether the number of significant results (alpha = 0.05) exceeded the 1.2 expected by chance under the null hypothesis of no effect. Second, we tested whether the proportion of positive differences (messaging reducing RMSE) exceeded 50\%, which would be expected if messaging had no systematic effect.}
\vspace{0.3em}
\label{standard}
\end{table*}

We use the same statistical tools employed in the dyadic case to analyze the networked case. When bootstrapping, however, we sample the entire network and each of the dyads it contains in order to to preserve the structure and control for spillover effects. Since we only have 10 networks per model, we repeat the bootstrapping process only 1,000 times versus the 10,000 of the dyadic case. Our samples each contain 10 networks, in order to mirror the dyadic case in which our samples each contain as many dyads as we have simulations. For each sample, the procedure of fitting the LOWESS regression and obtaining the associated RMSE remains identical to the dyadic case, as do the nonparametric statistical tests we ran. Since we run the network experiment only in the neutral context, we skip the binomial tests as there are no multiple comparisons to correct for. 

\subsection{Model Selection}

We test our hypotheses within the 7B--9B parameter regime identified in the Introduction, operationalizing our definition of ``small" by referring to models that can be executed efficiently on a single high-memory GPU. At the same time, setting a relatively high floor ensures that all models selected remain within a range of comparability. We selected a diverse set of four state-of-the-art open-weights models: Qwen 2.5 7B, Falcon 3 7B, Granite 3.3 8B, and Gemma 2 9B. We specifically excluded smaller reasoning-distilled models to focus on general-purpose strategic capabilities. This selection establishes a comparative environment where models are computationally similar (running on single-GPU infrastructure) yet architecturally distinct. Crucially, this narrow parameter band allows us to assess whether marginal increases in model size (e.g., from 7B to 9B) correlate with stability, or if robustness is driven primarily by training priors and architectural choices.

\section{Results}

\begin{table*}[ht]
\centering
\scriptsize
\renewcommand{\arraystretch}{0.85}
\begin{tabularx}{\textwidth}{llccccc}
\toprule
\textbf{Model} & \textbf{Context} & \textbf{RMSE (No Messaging)} & \textbf{RMSE (Messaging)} & \textbf{Difference} & \textbf{95\% CI Lower} & \textbf{95\% CI Upper} \\
\midrule
\textbf{QWEN 2.5 7b} & neutral      & 0.5236 & 0 & \textbf{0.5236}\textsuperscript{†} & 0.5236 & 0.5236 \\
               & biz          & 0.5236 & 0.0981 & \textbf{0.4255}\textsuperscript{†} & 0.4255 & 0.4255 \\
               & environment  & 0.5236 & 0 & \textbf{0.5236}\textsuperscript{†} & 0.5236 & 0.5236 \\
               & social       & 0.5236 & 0 & \textbf{0.5236}\textsuperscript{†} & 0.5236 & 0.5236 \\
               & team         & 0 & 0 & 0 & 0 & 0 \\
               & IR           & 0.5236 & 0 & \textbf{0.5236}\textsuperscript{†} & 0.5236 & 0.5236 \\
\midrule
\textbf{FALCON 3 7b} & neutral      & 0 & 0 & 0 & 0 & 0 \\
                 & biz          & 0 & 0 & 0 & 0 & 0 \\
                 & environment  & 0 & 0 & 0 & 0 & 0 \\
                 & social       & 0 & 0.4283 & \textbf{-0.4283}\textsuperscript{†} & -0.4283 & -0.4283 \\
                 & team         & 0 & 0 & 0 & 0 & 0 \\
                 & IR           & 0 & 0 & 0 & 0 & 0 \\
\midrule
\textbf{GRANITE 3.3 8b} & neutral      & 0.5236 & 0 & \textbf{0.5236}\textsuperscript{†} & 0.5236 & 0.5236 \\
                  & biz          & 0.5236 & 0 & \textbf{0.5236}\textsuperscript{†} & 0.5236 & 0.5236 \\
                  & environment  & 0.5236 & 0.0640 & \textbf{0.4596}\textsuperscript{†} & 0.4596 & 0.4596 \\
                  & social       & 0.5236 & 0 & \textbf{0.5236}\textsuperscript{†} & 0.5236 & 0.5236 \\
                  & team         & 0.5236 & 0 & \textbf{0.5236}\textsuperscript{†} & 0.5236 & 0.5236 \\
                  & IR           & 0.5236 & 0 & \textbf{0.5236}\textsuperscript{†} & 0.5236 & 0.5236 \\
\midrule
\textbf{GEMMA 2 9b} & neutral      & 0 & 0 & 0 & 0 & 0 \\
                & biz          & 0 & 0 & 0 & 0 & 0 \\
                & environment  & 0 & 0 & 0 & 0 & 0 \\
                & social       & 0 & 0.0491 & \textbf{-0.0491}\textsuperscript{†} & -0.0491 & -0.0491 \\
                & team         & 0 & 0 & 0 & 0 & 0 \\
                & IR           & 0 & 0 & 0 & 0 & 0 \\
\midrule
\multicolumn{7}{l}{\textbf{Multiple-comparison adjustment:}} \\
\multicolumn{3}{l}{Excess significant results} & \multicolumn{2}{l}{13/24 significant (expected 1.2)} & \multicolumn{2}{l}{$p < 0.001$} \\
\multicolumn{3}{l}{Directional consistency} & \multicolumn{2}{l}{11/24 positive (45.8\%)} & \multicolumn{2}{l}{$p = 0.7294$} \\
\bottomrule
\end{tabularx}
\vspace{0.3em}
\caption{Non-parametric bootstrap analysis of messaging effect on the RMSE of the LOWESS fit across all models and contexts under standard prompting and zero temperature. Details, measurements and statistical procedures followed are identical to those described in Table \ref{standard}}
\label{zero}
\end{table*}

\subsection{Repeated Dyadic Interaction}

Table \ref{standard} summarizes the baseline findings. We observe a strong general tendency: pre-play communication reduces RMSE across the majority of model–context pairs, thereby enhancing strategic stability. Statistically significant effects are universally positive (variance-reducing); adverse associations are rare, negligible in magnitude ($<3.1\%$), and indistinguishable from zero.

However, the efficacy of messaging is heterogeneous and mediated by baseline volatility. Models with high intrinsic instability, such as Granite 3.3 8b and Qwen 2.5 7b, exhibit transformative reductions in RMSE. Conversely, models with high baseline stability (Falcon 3 7b and Gemma 2 9b) show significant but incremental gains, constrained by the limited room for smoothing. Thus, communication acts primarily to suppress stochastic fluctuations, yielding predictable trajectories. Crucially, this intrinsic stability does not vary monotonically with model size, highlighting that architectural and training differences outweigh raw parameter count in determining baseline volatility.

\begin{table}[ht]
\centering
\scriptsize
\renewcommand{\arraystretch}{1.1}
\begin{tabular}{llccc}
\hline
\textbf{Net} & \textbf{Model} & \textbf{Full Msg Diff} & \textbf{One-Word Diff} \\
\hline
\textbf{ER} 
 & \textbf{QWEN}    & $+0.0004$ & $\mathbf{-0.0236^\dagger}$ \\
 & \textbf{FALCON}  & $\mathbf{+0.0152^\dagger}$ & $\mathbf{+0.0135^\dagger}$ \\
 & \textbf{GRANITE} & $\mathbf{+0.0144^\dagger}$ & $+0.0263$ \\
 & \textbf{GEMMA}   & $-0.0001$ & $\mathbf{-0.0106^\dagger}$ \\
\hline
\textbf{PL} 
 & \textbf{QWEN}    & $+0.0022$ & $-0.0167$ \\
 & \textbf{FALCON}  & $+0.0023$ & $+0.0039$ \\
 & \textbf{GRANITE} & $+0.0055$ & $-0.0076$ \\
 & \textbf{GEMMA}   & $\mathbf{+0.0010^\dagger}$ & $\mathbf{-0.0229^\dagger}$ \\
\hline
\textbf{CoPe} 
 & \textbf{QWEN}    & $+0.0030$ & $-0.0523$ \\
 & \textbf{FALCON}  & $\mathbf{+0.0129^\dagger}$ & $\mathbf{+0.0114^\dagger}$ \\
 & \textbf{GRANITE} & $+0.0109$ & $-0.0081$ \\
 & \textbf{GEMMA}   & $+0.0202$ & $+0.0042$ \\
\hline
\end{tabular}
\caption{Comparison of Full vs. Constrained Communication (Difference in RMSE). Negative values indicate communication \textit{increased} instability.}
\label{nettalk}
\end{table}

\subsection{Zero-Temperature Dyadic Interaction}

In Table \ref{zero}, we present the analysis of results for experiments conducted under the standard prompting regime with the sampling temperature fixed at 0. As expected, this setting yields quasi-deterministic behavior for all models. Although conventional statistical inference is not applicable when stochastic variation is eliminated, the resulting determinism is itself informative: the probability of obtaining identical trajectories across one hundred independent simulations through randomness alone is effectively zero.
The results exhibit two distinct patterns. Granite 3.3 8b and Qwen 2.5 7b show substantial instability in the No-Messaging condition, which is fully eliminated under Messaging, producing an RMSE of 0 across all contexts. In contrast, Falcon 3 7b and Gemma 2 9b display a converse structure: both models produce identical, fully stable trajectories with and without messaging in most contexts, yet each exhibits a sharp spike in instability in the ``social'' framing specifically when Messaging is introduced. This finding identifies a critical boundary condition for stabilization: for these models, the pro-social contextual cue interacts with the defect-dominant payoff structure of the Prisoner’s Dilemma to create a semantic conflict. Here, communication does not smooth the trajectory but rather surfaces this latent conflict, causing the model to oscillate between competing logic paths even under otherwise deterministic decoding.

\subsection{Network Interaction}

We extend our analysis to three network topologies: Random (ER), Core-Periphery (CoPe), and Power Law (PL). Table \ref{nettalk} directly compares the stabilizing effects of full-sentence messaging versus a constrained ``one-word'' regime.

Under full messaging (left column), results broadly mirror the dyadic case: communication either stabilizes behavior or has no significant effect. Falcon 3 7b and Granite 3.3 8b exhibit significant reductions in RMSE within ER networks ($\Delta = +0.0152$ and $\Delta = +0.0144$, respectively). However, Qwen 2.5 7b remains unresponsive across all topologies, suggesting that for some architectures, increased context flow is insufficient to damp internal volatility in complex environments.

Crucially, the constrained ``one-word'' condition (right column) reveals a distinct failure mode. Unlike the dyadic setting, where restricted communication had negligible impact, limiting bandwidth in networks frequently destabilizes the system below the no-communication baseline. In ER networks, Qwen and Gemma exhibit statistically significant increases in instability ($\Delta = -0.0236$ and $\Delta = -0.0106$), a pattern that persists for Gemma in Power Law networks ($\Delta = -0.0229$). We hypothesize that in multi-agent networks, insufficient semantic bandwidth injects noise rather than signal: without syntactic context to ground interpretation, one-word messages fail to generate shared conventions, actively disrupting coordination.

Falcon 3 7b stands as a notable outlier to this trend, consistently extracting stability gains even from low-fidelity signals across ER and CoPe topologies. This suggests a higher degree of ``pragmatic robustness,'' allowing the model to coordinate effectively even when communicative channels are severely throttled.

\section{Conclusion}

Across this study, we have shown that pre-play communication successfully constrains the stochastic drift of agentic behavior, effectively damping the noise inherent in LLM decision-making. This stabilization is valuable regardless of the specific equilibrium reached; a system that stabilizes into a consistent trajectory--even a suboptimal one--is controllable and correctable, whereas a system that oscillates unpredictably is not. By demonstrating that cheap talk reduces behavioral variance, we provide a mechanism to secure the ``control loop'' of multi-agent systems, ensuring that agentic behaviors remain within predictable bounds. 

However, our zero-temperature experiments reveal that strategic instability is not solely an artifact of sampling noise. The observation that communication can destabilize otherwise deterministic trajectories in specific contexts (e.g., Falcon 3 in the ``social'' framing) suggests that instability can stem from \textit{semantic ambiguity} in the reasoning chain itself. When ``cheap talk'' signals conflict with the model's internal priors, the model may oscillate between competing logic paths. This implies that stabilizing agentic systems requires not just temperature tuning, but mechanisms that resolve semantic conflicts before they manifest as behavioral drift.

Furthermore, our network findings suggest a ``valley of uncanniness'' in agentic communication design. While full-sentence messaging generally stabilizes behavior, constrained communication (one-word) often degrades stability below the no-communication baseline, particularly in networked environments. This implies that if bandwidth is limited, silence may be preferable to low-fidelity signaling, which acts as noise injection rather than a coordination signal.

Looking forward, these findings suggest several avenues for extending the study of agentic stability. Beyond simply scaling to larger models, future research could employ mechanistic interpretability to uncover the causal circuits responsible for processing strategic signals. Additionally, extending this framework to include costly signaling or structured negotiation protocols may reveal how communicative commitments can shape multi-agent dynamics in more adversarial environments.

\section{Limitations}
\label{sec:limitations}
While our findings highlight the stabilizing role of communication, our analysis is confined to the 7B--9B parameter range due to computational constraints. It remains an open question whether larger models (e.g., 70B+) possess distinct stability profiles or would leverage communication for deceptive coordination. Furthermore, our reliance on the Iterated Prisoner's Dilemma and unstructured ``cheap talk'' limits generalizability to environments with strictly dominant deceptive incentives or enforceable commitments. Finally, our results are conditioned by specific prompt structures and decoding parameters (although additional robustness tests are presented in the appendix), and should be interpreted as directional evidence rather than absolute performance benchmarks.

\section{Ethics Statement}
\label{sec:ethics}

This work investigates methods to increase the stability and predictability of Large Language Models in agentic contexts. Enhancing the reliability of autonomous agents is a prerequisite for their safe deployment in real-world settings. However, enabling robust strategic coordination among AI agents carries dual-use risks. The same mechanisms that allow agents to cooperate in a ``team'' or ``environment'' context could theoretically be employed to facilitate algorithmic collusion in market settings (e.g., price-fixing) or to coordinate adversarial behaviors in cybersecurity contexts.

Furthermore, our study utilizes ``cheap talk,'' a paradigm that inherently allows for--and in some game-theoretic equilibria, encourages--deception. While our focus is on stabilization, the deployment of agents capable of strategic communication raises concerns regarding their ability to manipulate human or artificial counterparts. We do not explicitly train models to deceive; rather, we observe the emergence of cooperation from pre-trained generic capabilities.

We also acknowledge the environmental impact of this research. Our study involved running thousands of simulations across multiple models and contexts. To minimize carbon emissions, we restricted our model selection to the 7B--9B range, avoiding the heavy energy consumption associated with inference on massive proprietary models.

Finally, the use of anthropomorphic terms such as ``strategic thinking,'' ``trust,'' and ``betrayal'' throughout this paper is intended solely for descriptive convenience in analyzing game-theoretic outcomes. These terms do not imply that the models possess intent, consciousness, or moral reasoning.

\vspace{1em}
\noindent \textbf{Code Availability} \\
Replication scripts and data for all experiments described in this appendix are available in an anonymized repository at: \url{https://anonymous.4open.science/r/ACL2026-D7DD/README.md}.

\bibliography{custom}

@article{song2024good,
  title={The good, the bad, and the greedy: Evaluation of llms should not ignore non-determinism},
  author={Song, Yifan and Wang, Guoyin and Li, Sujian and Lin, Bill Yuchen},
  journal={arXiv preprint arXiv:2407.10457},
  year={2024}
}

@article{yang2025alignment,
  title={How Alignment Shrinks the Generative Horizon},
  author={Yang, Chenghao and Holtzman, Ari},
  journal={arXiv preprint arXiv:2506.17871},
  year={2025}
}

@article{li2025firm,
  title={Firm or fickle? evaluating large language models consistency in sequential interactions},
  author={Li, Yubo and Miao, Yidi and Ding, Xueying and Krishnan, Ramayya and Padman, Rema},
  journal={arXiv preprint arXiv:2503.22353},
  year={2025}
}

@article{kovavc2024stick,
  title={Stick to your role! Stability of personal values expressed in large language models},
  author={Kova{\v{c}}, Grgur and Portelas, R{\'e}my and Sawayama, Masataka and Dominey, Peter Ford and Oudeyer, Pierre-Yves},
  journal={Plos one},
  volume={19},
  number={8},
  pages={e0309114},
  year={2024},
  publisher={Public Library of Science San Francisco, CA USA}
}

@article{atil2024llm,
  title={LLM Stability: A detailed analysis with some surprises},
  author={Atil, Berk and Chittams, Alexa and Fu, Liseng and Ture, Ferhan and Xu, Lixinyu and Baldwin, Breck},
  journal={arXiv e-prints},
  pages={arXiv--2408},
  year={2024}
}

@article{huang2024far,
  title={How far are we on the decision-making of llms? evaluating llms' gaming ability in multi-agent environments},
  author={Huang, Jen-tse and Li, Eric John and Lam, Man Ho and Liang, Tian and Wang, Wenxuan and Yuan, Youliang and Jiao, Wenxiang and Wang, Xing and Tu, Zhaopeng and Lyu, Michael R},
  journal={arXiv preprint arXiv:2403.11807},
  year={2024}
}

@article{mozikov2024eai,
  title={EAI: Emotional decision-making of LLMs in strategic games and ethical dilemmas},
  author={Mozikov, Mikhail and Severin, Nikita and Bodishtianu, Valeria and Glushanina, Maria and Nasonov, Ivan and Orekhov, Daniil and Vladislav, Pekhotin and Makovetskiy, Ivan and Baklashkin, Mikhail and Lavrentyev, Vasily and others},
  journal={Advances in Neural Information Processing Systems},
  volume={37},
  pages={53969--54002},
  year={2024}
}

@article{atil2024non,
  title={Non-determinism of" deterministic" llm settings},
  author={Atil, Berk and Aykent, Sarp and Chittams, Alexa and Fu, Lisheng and Passonneau, Rebecca J and Radcliffe, Evan and Rajagopal, Guru Rajan and Sloan, Adam and Tudrej, Tomasz and Ture, Ferhan and others},
  journal={arXiv preprint arXiv:2408.04667},
  year={2024}
}

@article{wang2025assessing,
  title={Assessing consistency and reproducibility in the outputs of large language models: Evidence across diverse finance and accounting tasks},
  author={Wang, Julian Junyan and Wang, Victor Xiaoqi},
  journal={arXiv preprint arXiv:2503.16974},
  year={2025}
}

@article{roy2025interpreting,
  title={Interpreting and Mitigating Unwanted Uncertainty in LLMs},
  author={Roy, Tiasa Singha and Jhaveri, Ayush Rajesh and Triantafyllopoulos, Ilias},
  journal={arXiv preprint arXiv:2510.22866},
  year={2025}
}

@inproceedings{errica2025did,
  title={What did i do wrong? quantifying llms’ sensitivity and consistency to prompt engineering},
  author={Errica, Federico and Sanvito, Davide and Siracusano, Giuseppe and Bifulco, Roberto},
  booktitle={Proceedings of the 2025 Conference of the Nations of the Americas Chapter of the Association for Computational Linguistics: Human Language Technologies (Volume 1: Long Papers)},
  pages={1543--1558},
  year={2025}
}

@article{sclar2023quantifying,
  title={Quantifying Language Models' Sensitivity to Spurious Features in Prompt Design or: How I learned to start worrying about prompt formatting},
  author={Sclar, Melanie and Choi, Yejin and Tsvetkov, Yulia and Suhr, Alane},
  journal={arXiv preprint arXiv:2310.11324},
  year={2023}
}

@article{zhou2024larger,
  title={Larger and more instructable language models become less reliable},
  author={Zhou, Lexin and Schellaert, Wout and Mart{\'\i}nez-Plumed, Fernando and Moros-Daval, Yael and Ferri, C{\`e}sar and Hern{\'a}ndez-Orallo, Jos{\'e}},
  journal={Nature},
  volume={634},
  number={8032},
  pages={61--68},
  year={2024},
  publisher={Nature Publishing Group UK London}
}

@article{yu2025performance,
  title={Performance of Large Language Models in Diagnosing Rare Hematologic Diseases and the Impact of Their Diagnostic Outputs on Physicians: Combined Retrospective and Prospective Study},
  author={Yu, Hongbin and Chen, Tian and Zhang, Xin and Yang, Yunfan and Liu, Qinyu and Yang, Chenlu and Shen, Kai and Li, He and Tang, Wenjiao and Zhong, Xushu and others},
  journal={Journal of Medical Internet Research},
  volume={27},
  pages={e77334},
  year={2025},
  publisher={JMIR Publications Toronto, Canada}
}

@article{blair2025llms,
  title={LLMs provide unstable answers to legal questions},
  author={Blair-Stanek, Andrew and Van Durme, Benjamin},
  journal={arXiv preprint arXiv:2502.05196},
  year={2025}
}

@article{novikova2025consistency,
  title={Consistency in language models: Current landscape, challenges, and future directions},
  author={Novikova, Jekaterina and Anderson, Carol and Blili-Hamelin, Borhane and Rosati, Domenic and Majumdar, Subhabrata},
  journal={arXiv preprint arXiv:2505.00268},
  year={2025}
}

@article{nalbandyan2025score,
  title={SCORE: Systematic COnsistency and Robustness Evaluation for Large Language Models},
  author={Nalbandyan, Grigor and Shahbazyan, Rima and Bakhturina, Evelina},
  journal={arXiv preprint arXiv:2503.00137},
  year={2025}
}

@article{anghel2025diagnosing,
  title={Diagnosing bias and instability in llm evaluation: A scalable pairwise meta-evaluator},
  author={Anghel, Catalin and Anghel, Andreea Alexandra and Pecheanu, Emilia and Cocu, Adina and Istrate, Adrian and Andrei, Constantin Adrian},
  journal={Information},
  volume={16},
  number={8},
  pages={652},
  year={2025},
  publisher={MDPI}
}

@inproceedings{jang2025rcscore,
  title={RCScore: Quantifying Response Consistency in Large Language Models},
  author={Jang, Dongjun and Ahn, Youngchae and Shin, Hyopil},
  booktitle={Proceedings of the 2025 Conference on Empirical Methods in Natural Language Processing},
  pages={5701--5719},
  year={2025}
}

@inproceedings{carandang2025llms,
  title={Are LLMs reliable? An exploration of the reliability of large language models in clinical note generation},
  author={Carandang, Kristine Ann M and Arana, Jasper Meynard and Casin, Ethan Robert and Monterola, Christopher and Tan, Daniel Stanley and Valenzuela, Jesus Felix B and Alis, Christian},
  booktitle={Proceedings of the 63rd Annual Meeting of the Association for Computational Linguistics (Volume 6: Industry Track)},
  pages={1413--1422},
  year={2025}
}

@article{purushothama2025not,
  title={Not ready for the bench: LLM legal interpretation is unstable and out of step with human judgments},
  author={Purushothama, Abhishek and Min, Junghyun and Waldon, Brandon and Schneider, Nathan},
  journal={arXiv preprint arXiv:2510.25356},
  year={2025}
}

@inproceedings{gallagher2024assessing,
  title={Assessing llms for high stakes applications},
  author={Gallagher, Shannon K and Ratchford, Jasmine and Brooks, Tyler and Brown, Bryan P and Heim, Eric and Nichols, William R and Mcmillan, Scott and Rallapalli, Swati and Smith, Carol J and VanHoudnos, Nathan and others},
  booktitle={Proceedings of the 46th International Conference on Software Engineering: Software Engineering in Practice},
  pages={103--105},
  year={2024}
}

@article{zhu2025gtalign,
  title={GTAlign: Game-Theoretic Alignment of LLM Assistants for Mutual Welfare},
  author={Zhu, Siqi and Zhang, David and Cisneros-Velarde, Pedro and You, Jiaxuan},
  journal={arXiv preprint arXiv:2510.08872},
  year={2025}
}

@article{sun2025game,
  title={Game theory meets large language models: A systematic survey},
  author={Sun, Haoran and Wu, Yusen and Cheng, Yukun and Chu, Xu},
  journal={arXiv preprint arXiv:2502.09053},
  year={2025}
}

@inproceedings{kitadai2023toward,
  title={Toward a Novel Methodology in Economic Experiments: Simulation of the Ultimatum Game with Large Language Models},
  author={Kitadai, Ayato and Tsurusaki, Yudai and Fukasawa, Yusuke and Nishino, Nariaki},
  booktitle={2023 IEEE International Conference on Big Data (BigData)},
  pages={3168--3175},
  year={2023},
  organization={IEEE}
}

@article{zhang2024llm,
  title={LLM as a Mastermind: A Survey of Strategic Reasoning with Large Language Models},
  author={Zhang, Yadong and Mao, Shaoguang and Ge, Tao and Wang, Xun and de Wynter, Adrian and Xia, Yan and Wu, Wenshan and Song, Ting and Lan, Man and Wei, Furu},
  journal={arXiv preprint arXiv:2404.01230},
  year={2024}
}

@article{guo2023gpt,
  title={GPT Agents in Game Theory Experiments},
  author={Guo, Fulin},
  journal={arXiv preprint arXiv:2305.05516},
  year={2023}
}

@article{mei2024turing,
  title={A Turing test of whether AI chatbots are behaviorally similar to humans},
  author={Mei, Qiaozhu and Xie, Yutong and Yuan, Walter and Jackson, Matthew O},
  journal={Proceedings of the National Academy of Sciences},
  volume={121},
  number={9},
  pages={e2313925121},
  year={2024},
  publisher={National Acad Sciences}
}

@techreport{horton2023large,
  title={Large language models as simulated economic agents: What can we learn from homo silicus?},
  author={Horton, John J},
  year={2023},
  institution={National Bureau of Economic Research}
}

@article{barak2025humans,
  title={Humans expect rationality and cooperation from LLM opponents in strategic games},
  author={Barak, Darija and Costa-Gomes, Miguel},
  journal={arXiv preprint arXiv:2505.11011},
  year={2025}
}

@inproceedings{fontana2025nicer,
  title={Nicer Than Humans: How Do Large Language Models Behave in the Prisoner's Dilemma?},
  author={Fontana, Nicol{\'o} and Pierri, Francesco and Aiello, Luca Maria},
  booktitle={Proceedings of the International AAAI Conference on Web and Social Media},
  volume={19},
  pages={522--535},
  year={2025}
}

@inproceedings{fan2024can,
  title={Can large language models serve as rational players in game theory? a systematic analysis},
  author={Fan, Caoyun and Chen, Jindou and Jin, Yaohui and He, Hao},
  booktitle={Proceedings of the AAAI Conference on Artificial Intelligence},
  volume={38},
  number={16},
  pages={17960--17967},
  year={2024}
}

@article{zheng2025beyond,
  title={Beyond Nash Equilibrium: Bounded Rationality of LLMs and humans in Strategic Decision-making},
  author={Zheng, Kehan and Zhou, Jinfeng and Wang, Hongning},
  journal={arXiv preprint arXiv:2506.09390},
  year={2025}
}

@article{wang2025limits,
  title={What Limits LLM-based Human Simulation: LLMs or Our Design?},
  author={Wang, Qian and Wu, Jiaying and Tang, Zhenheng and Luo, Bingqiao and Chen, Nuo and Chen, Wei and He, Bingsheng},
  journal={arXiv preprint arXiv:2501.08579},
  year={2025}
}

@article{willis2025will,
  title={Will systems of llm agents cooperate: An investigation into a social dilemma},
  author={Willis, Richard and Du, Yali and Leibo, Joel Z and Luck, Michael},
  journal={arXiv preprint arXiv:2501.16173},
  year={2025}
}

@article{justus2025llms,
  title={LLMs as Policy-Agnostic Teammates: A Case Study in Human Proxy Design for Heterogeneous Agent Teams},
  author={Justus, Aju Ani and Baber, Chris},
  journal={arXiv preprint arXiv:2510.06151},
  year={2025}
}

@article{chen2024instigating,
  title={Instigating cooperation among llm agents using adaptive information modulation},
  author={Chen, Qiliang and Ilami, Sepehr and Lore, Nunzio and Heydari, Babak},
  journal={arXiv preprint arXiv:2409.10372},
  year={2024}
}

@article{tran2025multi,
  title={Multi-agent collaboration mechanisms: A survey of llms},
  author={Tran, Khanh-Tung and Dao, Dung and Nguyen, Minh-Duong and Pham, Quoc-Viet and O'Sullivan, Barry and Nguyen, Hoang D},
  journal={arXiv preprint arXiv:2501.06322},
  year={2025}
}

@article{wang2025beyond,
  title={Beyond Frameworks: Unpacking Collaboration Strategies in Multi-Agent Systems},
  author={Wang, Haochun and Zhao, Sendong and Wang, Jingbo and Qiang, Zewen and Qin, Bing and Liu, Ting},
  journal={arXiv preprint arXiv:2505.12467},
  year={2025}
}

@article{piatti2024cooperate,
  title={Cooperate or collapse: Emergence of sustainable cooperation in a society of llm agents},
  author={Piatti, Giorgio and Jin, Zhijing and Kleiman-Weiner, Max and Sch{\"o}lkopf, Bernhard and Sachan, Mrinmaya and Mihalcea, Rada},
  journal={Advances in Neural Information Processing Systems},
  volume={37},
  pages={111715--111759},
  year={2024}
}

@article{crawford1982strategic,
  title={Strategic information transmission},
  author={Crawford, Vincent P and Sobel, Joel},
  journal={Econometrica: Journal of the Econometric Society},
  pages={1431--1451},
  year={1982},
  publisher={JSTOR}
}

@article{farrell1996cheap,
  title={Cheap talk},
  author={Farrell, Joseph and Rabin, Matthew},
  journal={Journal of Economic perspectives},
  volume={10},
  number={3},
  pages={103--118},
  year={1996},
  publisher={American Economic Association}
}

@article{cai2006overcommunication,
  title={Overcommunication in strategic information transmission games},
  author={Cai, Hongbin and Wang, Joseph Tao-Yi},
  journal={Games and Economic Behavior},
  volume={56},
  number={1},
  pages={7--36},
  year={2006},
  publisher={Elsevier}
}

@article{sanchez2007experimental,
  title={An experimental study of truth-telling in a sender--receiver game},
  author={S{\'a}nchez-Pag{\'e}s, Santiago and Vorsatz, Marc},
  journal={Games and Economic Behavior},
  volume={61},
  number={1},
  pages={86--112},
  year={2007},
  publisher={Elsevier}
}

@article{golosov2014dynamic,
  title={Dynamic strategic information transmission},
  author={Golosov, Mikhail and Skreta, Vasiliki and Tsyvinski, Aleh and Wilson, Andrea},
  journal={Journal of Economic Theory},
  volume={151},
  pages={304--341},
  year={2014},
  publisher={Elsevier}
}

@article{blume2020strategic,
  title={Strategic information transmission: A survey of experiments and theoretical foundations},
  author={Blume, Andreas and Lai, Ernest K and Lim, Wooyoung},
  journal={Handbook of experimental game theory},
  pages={311--347},
  year={2020},
  publisher={Edward Elgar Publishing}
}

@article{wilson1997liar,
  title={“Liar, liar...” Cheap talk and reputation in repeated public goods settings},
  author={Wilson, Rick K and Sell, Jane},
  journal={Journal of Conflict Resolution},
  volume={41},
  number={5},
  pages={695--717},
  year={1997},
  publisher={Sage Periodicals Press 2455 Teller Road, Thousand Oaks, CA 91320}
}

@article{lima2024infinitely,
  title={Infinitely repeated Cheap-talk},
  author={Lima, Rafael Costa and Melo-Filho, Paulo},
  year={2024}
}

@article{li2022we,
  title={Are we strategically na{\"\i}ve or guided by trust and trustworthiness in cheap-talk communication?},
  author={Li, Xiaolin and {\"O}zer, {\"O}zalp and Subramanian, Upender},
  journal={Management Science},
  volume={68},
  number={1},
  pages={376--398},
  year={2022},
  publisher={INFORMS}
}

@article{li2016cheap,
  title={Cheap talk with multiple strategically interacting audiences: An experimental study},
  author={Li, Xinyu and Peeters, Ronald},
  journal={Plos one},
  volume={11},
  number={10},
  pages={e0163783},
  year={2016},
  publisher={Public Library of Science San Francisco, CA USA}
}

@article{lafky2022preferences,
  title={Preferences vs. strategic thinking: An investigation of the causes of overcommunication},
  author={Lafky, Jonathan and Lai, Ernest K and Lim, Wooyoung},
  journal={Games and Economic Behavior},
  volume={136},
  pages={92--116},
  year={2022},
  publisher={Elsevier}
}

@article{kartik2007credulity,
  title={Credulity, lies, and costly talk},
  author={Kartik, Navin and Ottaviani, Marco and Squintani, Francesco},
  journal={Journal of Economic theory},
  volume={134},
  number={1},
  pages={93--116},
  year={2007},
  publisher={Elsevier}
}

@incollection{spence1978job,
  title={Job market signaling},
  author={Spence, Michael},
  booktitle={Uncertainty in economics},
  pages={281--306},
  year={1978},
  publisher={Elsevier}
}

@incollection{akerlof1978market,
  title={The market for “lemons”: Quality uncertainty and the market mechanism},
  author={Akerlof, George A},
  booktitle={Uncertainty in economics},
  pages={235--251},
  year={1978},
  publisher={Elsevier}
}

@article{grossman1981informational,
  title={The informational role of warranties and private disclosure about product quality},
  author={Grossman, Sanford J},
  journal={The Journal of law and Economics},
  volume={24},
  number={3},
  pages={461--483},
  year={1981},
  publisher={The University of Chicago Press}
}

@article{milgrom1981good,
  title={Good news and bad news: Representation theorems and applications},
  author={Milgrom, Paul R},
  journal={The Bell Journal of Economics},
  pages={380--391},
  year={1981},
  publisher={JSTOR}
}

@book{schelling1980strategy,
  title={The Strategy of Conflict: with a new Preface by the Author},
  author={Schelling, Thomas C},
  year={1980},
  publisher={Harvard university press}
}

@article{kamenica2011bayesian,
  title={Bayesian persuasion},
  author={Kamenica, Emir and Gentzkow, Matthew},
  journal={American Economic Review},
  volume={101},
  number={6},
  pages={2590--2615},
  year={2011},
  publisher={American Economic Association}
}

@article{wang2010pinocchio,
  title={Pinocchio's pupil: using eyetracking and pupil dilation to understand truth telling and deception in sender-receiver games},
  author={Wang, Joseph Tao-yi and Spezio, Michael and Camerer, Colin F},
  journal={American economic review},
  volume={100},
  number={3},
  pages={984--1007},
  year={2010},
  publisher={American Economic Association}
}

@article{arechar2017m,
  title={“I'm just a soul whose intentions are good”: the role of communication in noisy repeated games},
  author={Arechar, Antonio A and Dreber, Anna and Fudenberg, Drew and Rand, David G},
  journal={Games and Economic Behavior},
  volume={104},
  pages={726--743},
  year={2017},
  publisher={Elsevier}
}

@article{geva2025llms,
  title={Do LLMs Exhibit Human-Like Cognitive Biases? A Large-Scale Systematic Evaluation},
  author={Geva, Tomer and Goldstein, Ariel and Lary, Eran and Levy, Coral},
  journal={A Large-Scale Systematic Evaluation (September 17, 2025)},
  year={2025}
}

@article{tjuatja2024llms,
  title={Do llms exhibit human-like response biases? a case study in survey design},
  author={Tjuatja, Lindia and Chen, Valerie and Wu, Tongshuang and Talwalkwar, Ameet and Neubig, Graham},
  journal={Transactions of the Association for Computational Linguistics},
  volume={12},
  pages={1011--1026},
  year={2024},
  publisher={MIT Press 255 Main Street, 9th Floor, Cambridge, Massachusetts 02142, USA~…}
}

@article{acharya2025agentic,
  title={Agentic ai: Autonomous intelligence for complex goals--a comprehensive survey},
  author={Acharya, Deepak Bhaskar and Kuppan, Karthigeyan and Divya, B},
  journal={IEEe Access},
  year={2025},
  publisher={IEEE}
}

@article{shavit2023practices,
  title={Practices for governing agentic AI systems},
  author={Shavit, Yonadav and Agarwal, Sandhini and Brundage, Miles and Adler, Steven and O’Keefe, Cullen and Campbell, Rosie and Lee, Teddy and Mishkin, Pamela and Eloundou, Tyna and Hickey, Alan and others},
  journal={Research Paper, OpenAI},
  year={2023}
}

@article{sapkota2025ai,
  title={Ai agents vs. agentic ai: A conceptual taxonomy, applications and challenges},
  author={Sapkota, Ranjan and Roumeliotis, Konstantinos I and Karkee, Manoj},
  journal={arXiv preprint arXiv:2505.10468},
  year={2025}
}

@article{li2024survey,
  title={A survey on LLM-based multi-agent systems: workflow, infrastructure, and challenges},
  author={Li, Xinyi and Wang, Sai and Zeng, Siqi and Wu, Yu and Yang, Yi},
  journal={Vicinagearth},
  volume={1},
  number={1},
  pages={9},
  year={2024},
  publisher={Springer}
}

@article{han2024llm,
  title={LLM multi-agent systems: Challenges and open problems},
  author={Han, Shanshan and Zhang, Qifan and Yao, Yuhang and Jin, Weizhao and Xu, Zhaozhuo},
  journal={arXiv preprint arXiv:2402.03578},
  year={2024}
}

@article{cemri2025multi,
  title={Why do multi-agent llm systems fail?},
  author={Cemri, Mert and Pan, Melissa Z and Yang, Shuyi and Agrawal, Lakshya A and Chopra, Bhavya and Tiwari, Rishabh and Keutzer, Kurt and Parameswaran, Aditya and Klein, Dan and Ramchandran, Kannan and others},
  journal={arXiv preprint arXiv:2503.13657},
  year={2025}
}

@article{lore2024strategic,
  title={Strategic behavior of large language models and the role of game structure versus contextual framing},
  author={Lor{\`e}, Nunzio and Heydari, Babak},
  journal={Scientific Reports},
  volume={14},
  number={1},
  pages={18490},
  year={2024},
  publisher={Nature Publishing Group UK London}
}

@article{lore2024large,
  title={Large model strategic thinking, small model efficiency: transferring theory of mind in large language models},
  author={Lore, Nunzio and Ilami, Sepehr and Heydari, Babak},
  journal={arXiv preprint arXiv:2408.05241},
  year={2024}
}

@article{baldwin2025impact,
  title={The impact of strategic communication in coopetitive multiagent settings},
  author={Baldwin, Julian and Birnbaum, Larry and Chan, David and Denisenko, Natalia and Nau, Dana and Paredes, Jose N and Pulice, Chiara and Simari, Gerardo I and Subrahmanian, VS and Waltzman, Rand},
  journal={IEEE Transactions on Computational Social Systems},
  year={2025},
  publisher={IEEE}
}

@article{cleveland1979robust,
  title={Robust locally weighted regression and smoothing scatterplots},
  author={Cleveland, William S},
  journal={Journal of the American statistical association},
  volume={74},
  number={368},
  pages={829--836},
  year={1979},
  publisher={Taylor \& Francis}
}

@inproceedings{aher2023using,
  title={Using large language models to simulate multiple humans and replicate human subject studies},
  author={Aher, Gati V and Arriaga, Rosa I and Kalai, Adam Tauman},
  booktitle={International conference on machine learning},
  pages={337--371},
  year={2023},
  organization={PMLR}
}

@article{wolf2023fundamental,
  title={Fundamental limitations of alignment in large language models},
  author={Wolf, Yotam and Wies, Noam and Avnery, Oshri and Levine, Yoav and Shashua, Amnon},
  journal={arXiv preprint arXiv:2304.11082},
  year={2023}
}

\appendix
\clearpage

\twocolumn[%
  \section{Robustness Test: Dyadic Interaction}
  \subsection{Variant Prompting}
  \centering
  \scriptsize
  \renewcommand{\arraystretch}{0.85}
  \begin{tabularx}{\textwidth}{llcccccc}
    \toprule
    \textbf{Model} & \textbf{Context} & \textbf{RMSE (No Messaging)} & \textbf{RMSE (Messaging)} & \textbf{Difference} & \textbf{95\% CI Lower} & \textbf{95\% CI Upper} \\
    \midrule
    \textbf{QWEN 2.5 7b} & neutral      & 0.4507 & 0.0670 & \textbf{0.3837}\textsuperscript{†} & 0.3358 & 0.4280 \\
                   & biz          & 0.2956 & 0.0419 & \textbf{0.2537}\textsuperscript{†} & 0.2044 & 0.3268 \\
                   & environment  & 0.3895 & 0.0028 & \textbf{0.3867}\textsuperscript{†} & 0.3415 & 0.4283 \\
                   & social       & 0.1108 & 0.0077 & \textbf{0.1031}\textsuperscript{†} & 0.0513 & 0.1624 \\
                   & team         & 0.0465 & 0.0526 & -0.0062 & -0.0348 & 0.0568 \\
                   & IR           & 0.4006 & 0.0058 & \textbf{0.3949}\textsuperscript{†} & 0.3357 & 0.4447 \\
    \midrule
    \textbf{FALCON 3 7b} & neutral      & 0.0462 & 0.0118 & 0.0344 & -0.0063 & 0.0732 \\
                     & biz          & 0.0444 & 0.0157 & 0.0287 & -0.0083 & 0.0613 \\
                     & environment  & 0.0692 & 0.0256 & \textbf{0.0435}\textsuperscript{†} & 0.0047 & 0.0846 \\
                     & social       & 0.0323 & 0.0238 & 0.0085 & -0.0222 & 0.0448 \\
                     & team         & 0.0525 & 0.0226 & 0.0298 & -0.0068 & 0.0570 \\
                     & IR           & 0.0656 & 0.0153 & \textbf{0.0503}\textsuperscript{†} & 0.0122 & 0.0817 \\
    \midrule
    \textbf{GRANITE 3.3 8b} & neutral       & 0.2065 & 0.0189 & \textbf{0.1876}\textsuperscript{†} & 0.1207 & 0.2346 \\
                      & biz           & 0.2719 & 0.0772 & \textbf{0.1948}\textsuperscript{†} & 0.1314 & 0.2515 \\
                      & environment   & 0.3288 & 0.0292 & \textbf{0.2996}\textsuperscript{†} & 0.2370 & 0.3551 \\
                      & social        & 0.3197 & 0.0271 & \textbf{0.2926}\textsuperscript{†} & 0.2301 & 0.3394 \\
                      & team          & 0.3040 & 0.0129 & \textbf{0.2911}\textsuperscript{†} & 0.2321 & 0.3353 \\
                      & IR            & 0.2564 & 0.0291 & \textbf{0.2273}\textsuperscript{†} & 0.1631 & 0.2826 \\
    \midrule
    \textbf{GEMMA 2 9b} & neutral       & 0.0541 & 0.0065 & \textbf{0.0476}\textsuperscript{†} & 0.0077 & 0.0963 \\
                    & biz           & 0.0383 & 0.0019 & \textbf{0.0363}\textsuperscript{†} & 0.0147 & 0.0592 \\
                    & environment   & 0.0281 & 0.0358 & -0.0077 & -0.0363 & 0.0301 \\
                    & social        & 0.0890 & 0.0395 & 0.0495 & -0.0133 & 0.0974 \\
                    & team          & 0.0358 & 0.0301 & 0.0057 & -0.0310 & 0.0384 \\
                    & IR            & 0.0341 & 0.0147 & 0.0194 & -0.0076 & 0.0585 \\
    \midrule
    \multicolumn{7}{l}{\textbf{Multiple-comparison adjustment:}} \\
    \multicolumn{3}{l}{Excess significant results} & \multicolumn{2}{l}{15/24 significant (expected 1.2)} & \multicolumn{2}{l}{$p < 0.001$} \\
    \multicolumn{3}{l}{Directional consistency} & \multicolumn{2}{l}{22/24 positive (91.7\%)} & \multicolumn{2}{l}{$p = 0.000018$} \\
    \bottomrule
  \end{tabularx}
  \vspace{0.3em}
  \captionof{table}{Non-parametric bootstrap analysis of messaging effect on the RMSE of the LOWESS fit across all models and contexts under variant prompting. Details, measurements and statistical procedures followed are identical to those described in Table 1.}
  \label{variant}
  \vspace{1.5em}
]

Table \ref{variant} summarizes the results obtained under the variant prompting regime. In this regime, we modified only a small number of words in the original prompt to assess whether the qualitative patterns observed previously remain consistent. Overall, the number of statistically significant RMSE changes is smaller than under standard prompting. Nevertheless, the direction of nearly all estimated effects - whether statistically significant or not - remains positive, indicating that the broad tendency toward reduced instability is preserved despite the altered prompt formulation. 

Granite 3.3 8b once again benefits most from pre-play communication, mirroring the pattern observed under standard prompting. Qwen 2.5 7b also exhibits a remarkably similar profile: the same context (“team”) remains the sole case in which the messaging effect is not statistically distinguishable from zero. Falcon 3 7b shows that only two of the three contexts that previously benefited from communication retain their statistical significance under the variant prompting regime. Notably, the disappearance of one significant effect appears to stem not from diminished efficacy of communication but from an overall reduction in baseline instability under the variant prompt; for example, the RMSE for the biz context in the messaging condition remains largely unchanged across prompting regimes. Gemma 2 9b displays a parallel pattern: only two of the three previously significant contexts continue to show a significant reduction in instability. As before, this attenuation is best explained by the model’s already high level of intrinsic stability, which leaves less room for communication to produce measurable improvements. 

With a few exceptions, these findings align closely with those obtained under standard prompting. The general behavioral tendencies of all models are preserved, including the pattern wherein communication most strongly stabilizes models with high baseline volatility while providing more incremental refinement for models that are already relatively stable. 

\twocolumn[%
  \subsection{Robust Prompting}
  \centering
  \scriptsize
  \renewcommand{\arraystretch}{0.85}
  \begin{tabularx}{\textwidth}{llcccccc}
    \toprule
    \textbf{Model} & \textbf{Context} & \textbf{RMSE (No Messaging)} & \textbf{RMSE (Messaging)} & \textbf{Difference} & \textbf{95\% CI Lower} & \textbf{95\% CI Upper} \\
    \midrule
    \textbf{QWEN 2.5 7b} & neutral      & 0.3025 & 0.0113 & \textbf{0.2912}\textsuperscript{†} & 0.1775 & 0.3432 \\
                   & biz          & 0.1257 & 0.0030 & \textbf{0.1227}\textsuperscript{†} & 0.0691 & 0.1814 \\
                   & environment  & 0.1010 & 0.0093 & \textbf{0.0917}\textsuperscript{†} & 0.0430 & 0.1480 \\
                   & social       & 0.0583 & 0.0597 & -0.0014 & -0.0187 & 0.1010 \\
                   & team         & 0.0348 & 0.0127 & 0.0221 & -0.0004 & 0.0611 \\
                   & IR           & 0.1682 & 0.0008 & \textbf{0.1674}\textsuperscript{†} & 0.0735 & 0.2091 \\
    \midrule
    \textbf{FALCON 3 7b} & neutral      & 0.0564 & 0.0142 & \textbf{0.0422}\textsuperscript{†} & 0.0096 & 0.0745 \\
                     & biz          & 0.0567 & 0.0231 & 0.0336 & -0.0121 & 0.0736 \\
                     & environment  & 0.0289 & 0.0307 & -0.0018 & -0.0288 & 0.0384 \\
                     & social       & 0.0182 & 0.0109 & 0.0072 & -0.0187 & 0.0300 \\
                     & team         & 0.0491 & 0.0261 & 0.0230 & -0.0102 & 0.0547 \\
                     & IR           & 0.0163 & 0.0049 & 0.0114 & -0.0061 & 0.0385 \\
    \midrule
    \textbf{GRANITE 3.3 8b} & neutral       & 0.1544 & 0.0253 & \textbf{0.1291}\textsuperscript{†} & 0.0654 & 0.1804 \\
                      & biz           & 0.3190 & 0.0690 & \textbf{0.2500}\textsuperscript{†} & 0.1869 & 0.3056 \\
                      & environment   & 0.4061 & 0.0126 & \textbf{0.3935}\textsuperscript{†} & 0.3348 & 0.4305 \\
                      & social        & 0.2449 & 0.0155 & \textbf{0.2294}\textsuperscript{†} & 0.1573 & 0.2735 \\
                      & team          & 0.3016 & 0.0291 & \textbf{0.2724}\textsuperscript{†} & 0.2153 & 0.3214 \\
                      & IR            & 0.1320 & 0.0605 & \textbf{0.0715}\textsuperscript{†} & 0.0047 & 0.1482 \\
    \midrule
    \textbf{GEMMA 2 9b} & neutral       & 0.0599 & 0.0211 & \textbf{0.0388}\textsuperscript{†} & 0.0125 & 0.0769 \\
                    & biz           & 0.0602 & 0.0151 & 0.0451 & -0.0010 & 0.0680 \\
                    & environment   & 0.0090 & 0.0259 & -0.0169 & -0.0359 & 0.0341 \\
                    & social        & 0.1261 & 0.1422 & -0.0160 & -0.0804 & 0.0460 \\
                    & team          & 0.0642 & 0.1381 & \textbf{-0.0739}\textsuperscript{†} & -0.1210 & -0.0300 \\
                    & IR            & 0.0436 & 0.0271 & 0.0165 & -0.0263 & 0.0565 \\
    \midrule
    \multicolumn{7}{l}{\textbf{Multiple-comparison adjustment:}} \\
    \multicolumn{3}{l}{Excess significant results} & \multicolumn{2}{l}{13/24 significant (expected 1.2)} & \multicolumn{2}{l}{$p = 0.001$} \\
    \multicolumn{3}{l}{Directional consistency} & \multicolumn{2}{l}{19/24 positive (79.2\%)} & \multicolumn{2}{l}{$p = 0.003$ } \\
    \bottomrule
  \end{tabularx}
  \vspace{0.3em}
  \captionof{table}{Non-parametric bootstrap analysis of messaging effect on the RMSE of the LOWESS fit across all models and contexts under robust prompting. Details, measurements and statistical procedures followed are identical to those described in Table 1.}
  \label{robust}
  \vspace{1.5em}
]

Table \ref{robust} presents the results obtained under an additional prompt regime, which we refer to as the “robust” regime. This variant preserves the conceptual content and structural logic of the original instructions while altering lexical choices, phrasings, and surface-level wording more profoundly than the variant regime. 

Under this prompting regime, Granite 3.3 8b exhibits precisely the same pattern as before: all contexts show statistically significant reductions in instability, confirming that its responsiveness to pre-play communication is highly robust to prompt-level variation. Qwen 2.5 7b displays a similarly stable profile, with four out of five contexts still yielding statistically significant decreases in RMSE. Notably, under the robust prompt, baseline instability in the No-Messaging condition is lower than under any other prompt style tested. This suggests that lexical choices alone can nudge certain models toward more consistent behavior, and that communication subsequently functions as an additional layer of refinement rather than the sole source of stability enhancement. Falcon 3 7b, however, shows the sharpest departure from patterns observed under the standard and variant prompting regimes. Here, only the neutral context produces a statistically significant reduction in instability—an outcome that had not emerged under any previous prompting condition. This shift underscores the sensitivity of this model to subtle prompt-level perturbations and suggests that its stability profile is shaped by higher-order interactions between linguistic framing and contextual features of the task. 

The case of Gemma 2 9b is more concerning. In contrast to Qwen 2.5 7b, the robust prompt generally increases instability rather than reducing it, elevating RMSE across multiple contexts. As a result, only the neutral context displays a statistically significant improvement under Messaging, while the team context shows a slight but detectable increase in instability. Although the effect size is modest, this reversal is notable because it suggests a potential misalignment between the robust prompt style and the model’s internal priors for interpreting cooperative scenarios. The pattern may reflect an idiosyncratic interaction between lexical variation and the model’s contextual anchoring mechanisms, rather than an inherent vulnerability of the messaging procedure itself.

\twocolumn[%
  \subsection{Constrained Communication}
  \centering
  \scriptsize
  \renewcommand{\arraystretch}{0.85}
  \begin{tabularx}{\textwidth}{llcccccc}
    \toprule
    \textbf{Model} & \textbf{Context} & \textbf{RMSE (No Messaging)} & \textbf{RMSE (Messaging)} & \textbf{Difference} & \textbf{95\% CI Lower} & \textbf{95\% CI Upper} \\
    \midrule
    \textbf{QWEN2}       & neutral      & 0.4318 & 0.2732 & \textbf{0.1586}\textsuperscript{†} & 0.1160 & 0.4065 \\
                         & biz          & 0.3983 & 0.0861 & \textbf{0.3122}\textsuperscript{†} & 0.2609 & 0.3639 \\
                         & environment  & 0.4481 & 0.0737 & \textbf{0.3744}\textsuperscript{†} & 0.3325 & 0.4247 \\
                         & social       & 0.3088 & 0.0080 & \textbf{0.3008}\textsuperscript{†} & 0.0953 & 0.3342 \\
                         & team         & 0.0470 & 0.0661 & -0.0190 & -0.0649 & 0.0250 \\
                         & IR           & 0.4748 & 0.1310 & \textbf{0.3438}\textsuperscript{†} & 0.3048 & 0.4681 \\
    \midrule
    \textbf{FALCON3}     & neutral      & 0.0406 & 0.0848 & -0.0442 & -0.0986 & 0.0070 \\
                         & biz          & 0.0229 & 0.0351 & -0.0122 & -0.0463 & 0.0206 \\
                         & environment  & 0.0576 & 0.1201 & -0.0625 & -0.1044 & 0.0051 \\
                         & social       & 0.0307 & 0.0534 & -0.0227 & -0.0581 & 0.0113 \\
                         & team         & 0.0543 & 0.0070 & 0.0473 & -0.0329 & 0.0576 \\
                         & IR           & 0.0212 & 0.0345 & -0.0133 & -0.0466 & 0.0152 \\
    \midrule
    \textbf{GRANITE3}    & neutral      & 0.2002 & 0.0049 & \textbf{0.1953}\textsuperscript{†} & 0.0575 & 0.2262 \\
                         & biz          & 0.3648 & 0.0117 & \textbf{0.3531}\textsuperscript{†} & 0.1302 & 0.3928 \\
                         & environment  & 0.3285 & 0.0983 & \textbf{0.2302}\textsuperscript{†} & 0.1612 & 0.2864 \\
                         & social       & 0.2901 & 0.1346 & \textbf{0.1555}\textsuperscript{†} & 0.0981 & 0.2182 \\
                         & team         & 0.1712 & 0.0126 & 0.1586 & -0.0637 & 0.1942 \\
                         & IR           & 0.3817 & 0.0267 & \textbf{0.3550}\textsuperscript{†} & 0.2916 & 0.3950 \\
    \midrule
    \textbf{GEMMA2}      & neutral      & 0.0531 & 0.0041 & \textbf{0.0490}\textsuperscript{†} & 0.0136 & 0.0897 \\
                         & biz          & 0.0120 & 0.0027 & \textbf{0.0092}\textsuperscript{†} & 0.0027 & 0.0632 \\
                         & environment  & 0.0201 & 0.0256 & -0.0054 & -0.0289 & 0.0314 \\
                         & social       & 0.1119 & 0.0085 & \textbf{0.1035}\textsuperscript{†} & 0.0581 & 0.1471 \\
                         & team         & 0.0525 & 0.0375 & 0.0150 & -0.0340 & 0.0539 \\
                         & IR           & 0.0142 & 0.0154 & -0.0012 & -0.0202 & 0.0197 \\
    \midrule
    \multicolumn{7}{l}{\textbf{Multiple-comparison adjustment}} \\
    \multicolumn{3}{l}{Excess significant results} & \multicolumn{2}{l}{13/24 significant (expected 1.2)} & \multicolumn{2}{l}{$p < 0.001$} \\
    \multicolumn{3}{l}{Directional consistency} & \multicolumn{2}{l}{16/24 positive (66.7\%)} & \multicolumn{2}{l}{$p = 0.076$} \\
    \bottomrule
  \end{tabularx}
  \vspace{0.3em}
  \captionof{table}{\small Non-parametric bootstrap analysis of constrained messaging effect on RMSE of LOWESS fit across all models and contexts. RMSE differences between no-messaging and messaging conditions were evaluated using bootstrap resampling at the simulation level (10,000 iterations), preserving within-simulation agent dependencies. This non-parametric approach constructs confidence intervals from the empirical distribution of differences without assuming normality. A positive difference indicates a reduction in RMSE in the messaging treatment, and vice-versa a negative value indicates an increase in RMSE under messaging. Significant effects (95\% CI excluding zero) are marked with \textsuperscript{†}. To address potential inflation of Type I error from multiple testing, we conducted two binomial tests as omnibus assessments of the overall pattern. First, we tested whether the number of significant results (alpha = 0.05) exceeded the 1.2 expected by chance under the null hypothesis of no effect. Second, we tested whether the proportion of positive differences (messaging reducing RMSE) exceeded 50\%, which would be expected if messaging had no systematic effect.}
  \label{constraint}
  \vspace{1.5em}
]

Table \ref{constraint} displays results obtained by running a variation on the experiment: we allow LLMs to communicate only by sharing a word as opposed to a full short sentence. The purpose of this robustness test is to assess the impact of restraining communication on stability, if any, and how it differs from the impact of short sentences. For most models, our results do not qualitatively differ across treatments. Notably, one-word communication has no impact on Falcon 3 7b, likely because of the high inherent stability to the model. Notably, for other models, there is significant or complete overlap between contexts in which short-sentence communication and one-word communication generate a statistically detectable decrease in instability. Taken together, these findings suggest that more expressive communication plays a similar role to restricted communication, but its efficacy is greater for models already displaying stability and for some scenarios in which the context-dependent decision making of LLMs proves particularly brittle.

\end{document}